\documentclass[journal]{IEEEtran}
\usepackage{amsmath,amssymb,amsfonts}
\usepackage{algorithmic}
\usepackage{graphicx}
\usepackage{textcomp}
\usepackage{xcolor}
\usepackage{array}
\usepackage{booktabs} % 表格修改
\usepackage[ruled, linesnumbered, vlined]{algorithm2e}
\usepackage{textcomp}
\usepackage{multirow}
\usepackage{bm}
\usepackage{booktabs}
\usepackage{ntheorem}
\usepackage{subfigure} 
\usepackage{orcidlink} 
\hypersetup{hidelinks}
\usepackage{mathrsfs}
\usepackage{mathtools}
\usepackage{makecell}
\usepackage{enumitem} % 确保加载 enumitem 包
\usepackage{epstopdf}
\usepackage{cases}
\usepackage{hyperref}
\usepackage{booktabs} % 三线表支持
\usepackage{multirow}
\usepackage{tabularx}
\definecolor{color}{rgb}{0, 0, 0}
\definecolor{color_red}{rgb}{1, 0, 0}
\definecolor{color_revise}{rgb}{0, 0, 0}

\def\BibTeX{{\rm B\kern-.05em{\sc i\kern-.025em b}\kern-.08em
		T\kern-.1667em\lower.7ex\hbox{E}\kern-.125emX}}
\usepackage{cases}
\theoremseparator{.}

\theorembodyfont{}

\begin{document}

\title{Generative Artificial Intelligence for Beamforming in Low-Altitude Economy}

\author{
Geng Sun$^{\orcidlink{0000-0001-7802-4908}}$,
Jia Qi,
Chuang Zhang$^{\orcidlink{0000-0002-0505-0512}}$,
Xuejie Liu,
Jiacheng Wang$^{\orcidlink{0000-0003-1252-8761}}$,
Dusit~Niyato$^{\orcidlink{0000-0002-7442-7416}}$,~\IEEEmembership{Fellow,~IEEE,}\\
Yuanwei Liu$^{\orcidlink{0000-0002-6389-8941}}$,~\IEEEmembership{Fellow,~IEEE,}
and Dong In Kim$^{\orcidlink{0000-0001-7711-8072}}$,~\IEEEmembership{Fellow,~IEEE}

\thanks{Geng Sun is with the College of Computer Science and Technology, Key Laboratory of Symbolic Computation and Knowledge Engineering of Ministry of Education, Jilin University, Changchun 130012, China, and also with the College of Computing and Data Science, Nanyang Technological University, Singapore 639798 (e-mail: sungeng@jlu.edu.cn).}
\thanks{Jia Qi and Xuejie Liu are with the College of Computer Science and Technology, Jilin University, Changchun 130012, China (emails: qijia24@mails.jlu.edu.cn, xuejie@jlu.edu.cn)}
\thanks{Chuang Zhang is with the College of Computer Science and Technology, Key Laboratory of Symbolic Computation and Knowledge Engineering of Ministry of Education, Jilin University, Changchun 130012, China, and also with the Singapore University of Technology and Design, Singapore 487372 (email: chuangzhang1999@gmail.com).
}
\thanks{Jiacheng Wang and Dusit Niyato are with the College of Computing and Data Science, Nanyang Technological University, Singapore 639798 (emails: jiacheng.wang@ntu.edu.sg, dniyato@ntu.edu.sg).}
\thanks{Yuanwei Liu is with the Department of Electrical and Electronic Engineering, The University of Hong Kong, Hong Kong, China (e-mail: yuanwei@hku.hk).}
\thanks{Dong In Kim is with the Department of Electrical and Computer Engineering, Sungkyunkwan University, Suwon 16419, South Korea (e-mail: dongin@skku.edu).}
\thanks{\textit{(Corresponding author: Xuejie Liu.)}}
}

% The paper headers
\markboth{Journal of \LaTeX\ Class Files,~Vol.~14, No.~8, August~2015}%
{Shell \MakeLowercase{\textit{et al.}}: Bare Demo of IEEEtran.cls for Computer Society Journals}

\IEEEtitleabstractindextext{
    \begin{abstract}	
The growth of low-altitude economy (LAE) has driven a rising demand for efficient and secure communications. However, conventional beamforming optimization techniques struggle in the complex LAE environments. In this context, generative artificial intelligence (GenAI) methods provide a promising solution. In this article, we first introduce the core concepts of LAE and the roles of beamforming in advanced communication technologies for LAE. We then examine their interrelation, followed by an analysis of the limitations of conventional beamforming methods. Next, we provide an overview of how GenAI methods enhance the process of beamforming with a focus on its applications in LAE. Furthermore, we present a case study using a generative diffusion model (GDM)-based algorithm to enhance the performance of collaborative beamforming-enabled secure communications in LAE and simulation results demonstrate that our approach achieves 23\% improvement in terms of secrecy rate and 18\% reduction in energy consumption compared to baseline algorithms. Finally, promising research opportunities are identified.  
	\end{abstract}
	
}

\maketitle
\IEEEdisplaynontitleabstractindextext
\IEEEpeerreviewmaketitle

%	
% Introduction
%

\section{Introduction}
\label{Sec:Introduction}

\par Low-altitude economy (LAE) is an economic model that relies on the utilization of low-altitude airspace for various commercial activities, with widespread applications in fields such as precision agriculture, urban logistics and disaster rescue \cite{jiang20236g}. By enabling efficient aerial resource allocation, LAE can quickly adapt to the demands of high-density users and dynamic tasks. Moreover, LAE enhances the service response speed and coverage by enabling rapid resource deployment, especially in remote or congested areas where traditional infrastructure may be limited or unavailable. However, to meet these demands, the communication system of LAE faces challenging requirements for supporting efficient and flexible resource management in rapidly changing environments \cite{Yuan2025Groundtosky}.

\par As a core technology in wireless communication systems, beamforming has been widely applied in LAE networking. By controlling the phase and amplitude of antenna arrays, beamforming enhances signal directionality and strength, thereby addressing the challenges such as signal coverage, interference suppression and spectrum allocation. However, conventional beamforming optimization methods face limitations in LAE scenarios, which involve dynamic channel conditions, high-density interference and complex task requirements. These methods struggle to adapt to real-time environmental changes and lack sufficient flexibility for multi-user collaboration and resource allocation. Therefore, relying solely on conventional beamforming optimization methods is insufficient to fully meet the communication demands of LAE networking.

\par Fortunately, recent advances in artificial intelligence (AI) technologies offer promising solutions to these issues. Traditional AI (TAI) techniques are built on supervised learning, directly mapping labeled channel data to optimal beamforming parameters. Moreover, TAI-based methods utilize deterministic algorithms with reduced parameter dimensionality and memory footprint. As a result, they are well suited for the lightweight deployment, which requires minimal computational resources for implementation on resource-constrained low-altitude platforms (LAPs). However, the reliance of TAI on predefined mappings and labeled datasets limits its generalization and adaptability in highly dynamic LAE environments with non-stationary channels or sparse data. 

\par In contrast, generative AI (GenAI) models can inherently address these challenges. Specifically, GenAI models use self-supervised learning that enables adaptation to diverse environments by capturing stochastic propagation phenomena and transferring knowledge under varied conditions, thus providing robust performance without environment-specific tuning. Furthermore, GenAI can synthesize realistic channel realizations and adapt to new situations with minimal training data. Additionally, uncertainty quantification mechanisms in generative models support optimal decision-making under limited channel state information, effectively balancing performance with data constraints in LAE communication scenarios.

\begin{figure*}[!hbt] 
	\centering
	\includegraphics[width =\textwidth]{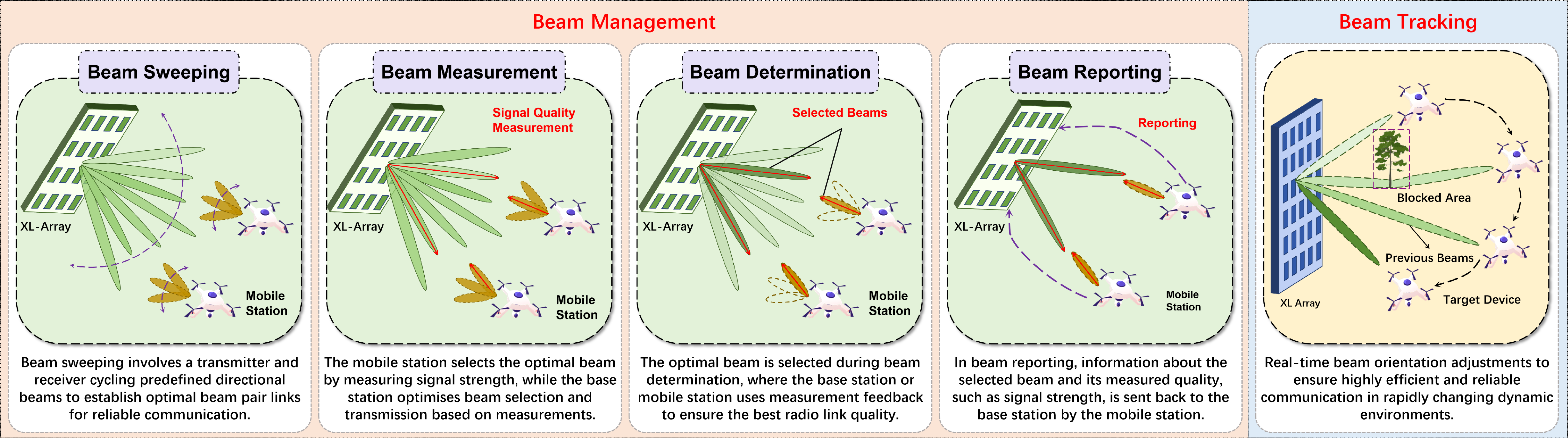}
	\caption{This figure presents the two key technologies in beamforming: beam management and beam tracking, each of which illustrates how it contributes to optimizing communication performance.}
	\label{fig_timeline}
\end{figure*}

\par By applying generative models, GenAI methods can capture complex channel characteristics and generate highly accurate channel state information (CSI), thereby improving the design and adjustment strategies for beamforming. Moreover, GenAI models present potential for adapting to frequent channel variations in dynamic environments and enhancing resource utilization and interference management. Crucially, GenAI models can learn from diverse data distributions, which enables them to evolve given various communication requirements within LAE.

\par Motivated by the potential of GenAI models in addressing these challenges, we focus on integrating them with beamforming within the LAE. The main contributions of this work are as follows:

\begin{itemize}
    \item
    We first introduce the characteristics of LAE and the fundamental concepts of beamforming. Then, we elaborate on the applications of beamforming in both communications and sensing in the context of LAE. Finally, we discuss the challenges faced by beamforming in LAE scenarios.
    \item 
    We describe several core techniques and applications of GenAI. Subsequently, we discuss how these GenAI techniques can be applied to beam management and beam tracking, as well as their specific roles in beamforming optimization for LAE.
    \item 
    We propose a framework based on GenAI to optimize the performance of collaborative beamforming-enabled secure communications in LAE. Simulation results show that the proposed framework significantly improves the secrecy rate and energy consumption of the system.
\end{itemize}

\section{Overview of Low-altitude Economy Beamforming}
\label{Sec:Overview of Low-altitude  Economy Beamforming}

\par In this section, we first introduce the core concepts of LAE. Next, we outline the fundamental processes of beamforming. Finally, we explore the various applications of beamforming in LAE and emphasize its roles in communications and sensing.

\subsection{Low-Altitude Economy}
\par LAE is an emerging economic model that uses low-altitude airspace to tackle the challenges of modern cities. By incorporating advanced communication technologies and intelligent computing into LAPs, LAE improves the efficiency and responsiveness of industries such as logistics, urban mobility, and emergency services. The key characteristics and challenges of LAE are as follows.

\subsubsection{Heterogeneous Communication Demands} The diverse operational requirements of LAPs in LAE scenarios create heterogeneous communication demands, thus bringing significant challenges in communication resource allocation. Specifically, in disaster reconnaissance and urban surveillance scenarios, LAPs must support high data rates to enable real-time video transmission \cite{Jiang2025integrated}. In comparison, applications such as urban air taxis rely on ultra-reliable links with extremely low latency since both reliability and responsiveness are critical for ensuring passenger safety. Meanwhile, for agricultural protection, LAPs prioritize wide coverage over transmission rate. Consequently, these varied and often conflicting requirements necessitate advanced communication solutions capable of dynamic resource allocation.

\subsubsection{Dynamic Network Topologies} 
The mission-driven mobility of LAPs creates rapidly shifting network structures that challenge conventional communication paradigms. Specifically, large-scale agricultural spraying operations need real-time topology adaptation as LAP swarms cover vast farmlands \cite{Wei2025multiuav}. Moreover, emergency response like wildfire monitoring requires LAPs to dynamically form temporary relay chains for extended coverage in disaster zones. In addition, for urban air logistics, delivery LAP fleets execute sub-second topology switches for collision avoidance in congested corridors. Consequently, mission scenarios involving highly dynamic LAP mobility demand self-organizing networks capable of seamless topology transitions under strict latency constraints.

\subsubsection{Complex Communication Environments} LAP operations face severe wireless signal disruption. Specifically, urban package delivery encounters unpredictable obstructions and reflections from buildings and alleys. Moreover, in precision crop dusting over terraced fields, agricultural LAPs suffer signal degradation and physical obstructions from dense foliage. Furthermore, event monitoring over crowded areas experiences intense interference from numerous nearby LAPs, phones and other wireless devices. As a result, these environmental and interference challenges necessitate robust communication strategies adaptable to rapidly changing signal conditions for reliable and stable system performance.

\par Given the aforementioned characteristics and challenges of LAE, effective wireless signal management and optimization in low-altitude operations are crucial. In this context, beamforming emerges as a relevant solution, which has substantial potential to tackle signal-related challenges.

\subsection{Beamforming} 

\par Beamforming is a signal processing technique for optimizing the directionality of electromagnetic waves by controlling the phase and amplitude of each antenna element in an antenna array. As shown in Fig.~\ref{fig_timeline}, the beamforming-related technology details are discussed below.

\subsubsection{Beam Management}
\par Beam management focuses on acquiring and maintaining reliable beam pairs to ensure robust directional communications. 

\begin{itemize}
    \item \textbf{Beam Sweeping:} 
    Beam sweeping involves scanning predefined beam directions to identify optimal signal paths. 

    \item \textbf{Beam Measurement:}
    Beam measurement evaluates signal strength and interference across all candidate beams. 

    \item \textbf{Beam Determination:}
    Based on the results of beam measurements, the beam determination algorithm selects the optimal configuration.

    \item \textbf{Beam Reporting:}
    Beam reporting involves transmitting feedback on beam status to network controllers or centralized systems. 
\end{itemize}

\subsubsection{Beam Tracking}
\par Beam tracking continuously updates the beam direction in response to target movement, which ensures that the beam remains accurately aligned throughout the transmission process.

\subsection{Beamforming Solutions in Low-Altitude Economy}

\subsubsection{Beamforming for Heterogeneous Communication Demands} Beamforming offers an effective means of addressing heterogeneous communication demands in LAE scenarios by directing signal energy toward specific spatial directions. For instance, for high data rate applications, beamforming enhances the throughput of data-intensive transmissions by focusing strong beams on video-streaming LAPs. Conversely, in ultra-reliable communication scenarios, beamforming helps establish robust links for safety-critical platforms by precisely steering energy and suppressing potential interference. Furthermore, for wide-area coverage, beamforming extends communication range by targeting distant agricultural LAPs. Therefore, this spatial flexibility enables adaptive resource allocation tailored to the specific communication needs of each LAP, thereby facilitating the efficient coordination of diverse demands across the network.

\subsubsection{Beamforming for Dynamic Network Topologies} Beamforming effectively addresses the challenge of constantly changing network connections in LAE by enabling rapid redirection of communication signals. For example, in scenarios such as multi-LAP emergency response operations, beamforming allows communication nodes to quickly shift their focused signal beams as LAPs relocate or reconfigure formations. Similarly, for large-scale LAP-based delivery systems, beamforming enables central communication hubs to dynamically track and maintain robust signal links with other LAPs during critical maneuvers such as takeoff, landing and obstacle navigation. As a result, by continuously adjusting the beam direction towards moving targets, beamforming significantly improves connection stability and minimizes disruptions in these highly dynamic environments.

\subsubsection{Beamforming for Complex Communication Environments} Beamforming provides a powerful solution for overcoming signal degradation in challenging low-altitude settings by concentrating signal energy along desired paths. Specifically, for LAPs operating in urban areas, beamforming helps overcome signal blockage caused by buildings by focusing the transmission towards the intended receiver, thereby finding stronger paths through or around obstacles, which improves link reliability. Moreover, in agricultural applications, beamforming effectively counters the signal weakening and physical obstruction caused by dense crops by directing a focused beam, thus ensuring stable control and data transfer over longer distances. Furthermore, in crowded event areas with many LAPs and devices, beamforming minimizes interference by precisely directing signals towards intended receivers. Consequently, this focused approach is essential for maintaining clear communication in congested and cluttered environments.

\subsection{Applications of Beamforming in Low-Altitude Economy}
\begin{figure*}[!hbt] 
	\centering
	\includegraphics[width =\textwidth]{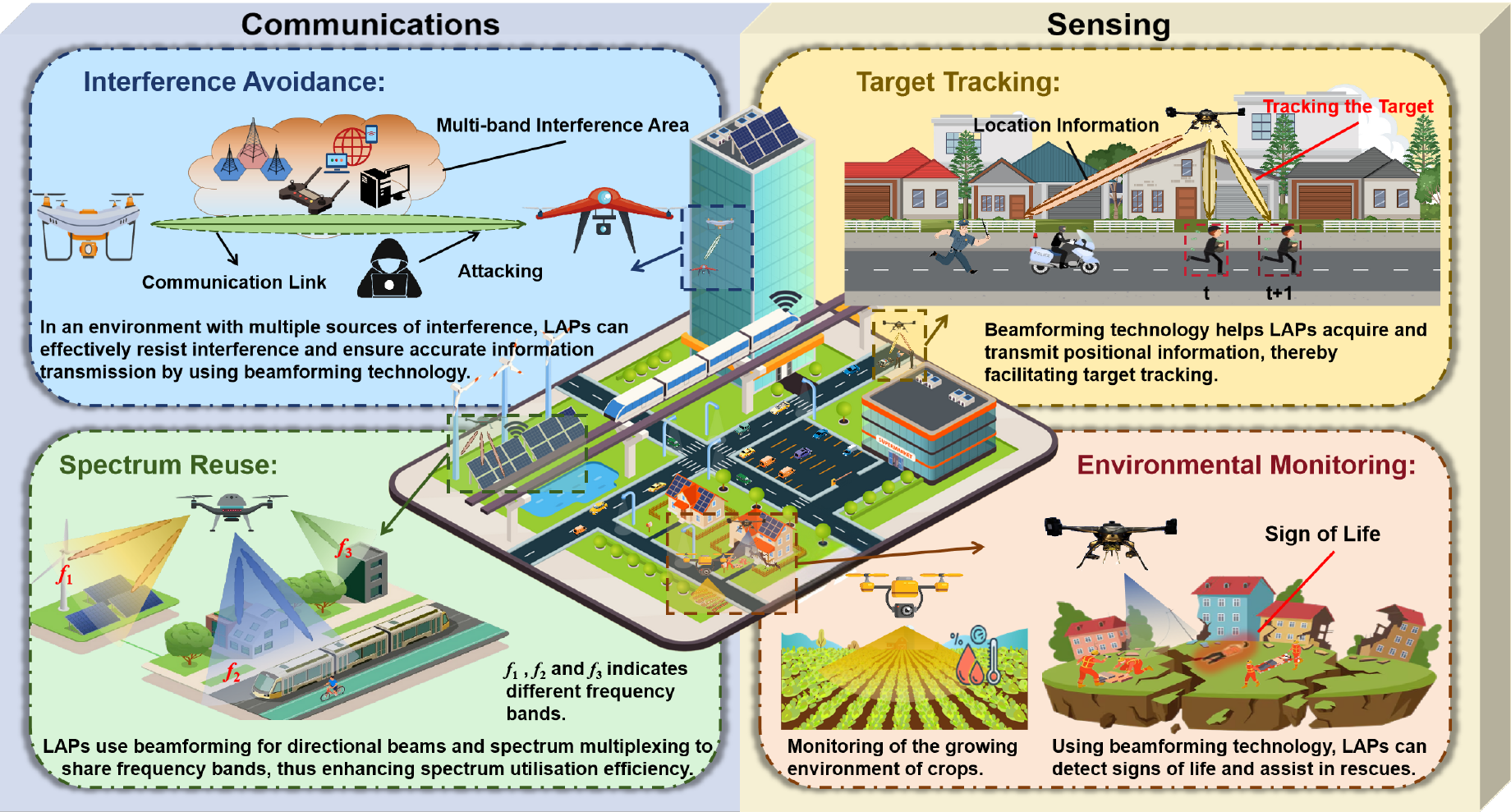}
	\caption{Beamforming enables efficient communications and sensing applications in the LAE, such as interference avoidance, spectrum reuse, target tracking and environmental monitoring.}
	\label{fig_Applications_of_beamforming_in_LAE}
\end{figure*}

\begin{figure*}[!hbt]
        \centering
        \includegraphics[width =\textwidth]{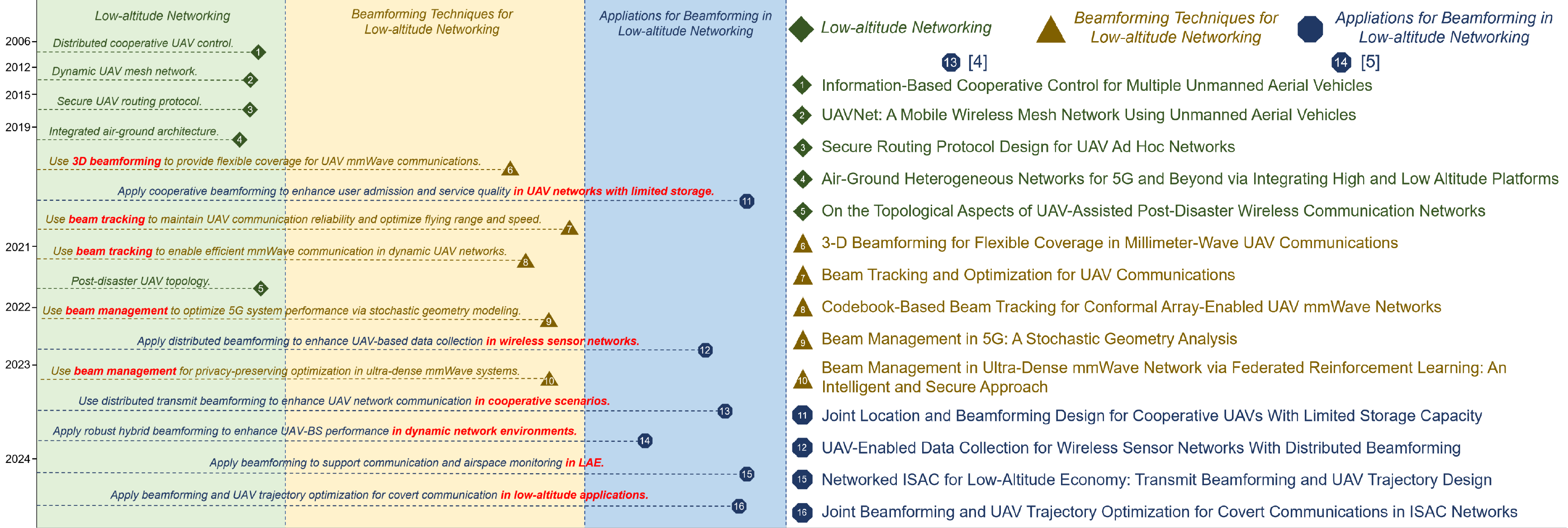}
        \caption{An overview of the timeline presenting the developments in low-altitude networking, beamforming techniques, and applications of beamforming in low-altitude networking.}   
        \label{fig_literature_review}
\end{figure*}

Beamforming technology supports a wide range of applications in low-altitude networking, broadly classified into communication and sensing domains. Specifically, Fig. \ref{fig_Applications_of_beamforming_in_LAE} highlights the classical applications within these domains. Meanwhile, a comprehensive literature review diagram is presented in Fig. \ref{fig_literature_review}, which offers an overview of the low-altitude networking, key beamforming techniques, and their applications in low-altitude networking.

\subsubsection{Communications}
In the rapidly evolving landscape of LAE applications, beamforming technology can address the unique communication challenges encountered by aerial networks. First, LAE networks frequently contend with severe interference in dense urban environments where multiple signal sources compete for limited spectrum resources~\cite{hanna2022distributed}. To address the challenge, beamforming effectively reduces interference by directing signal energy more accurately toward a specified target through the enhanced directivity of the antenna array. Therefore, such directional energy transmission reduces signal spread and minimizes interference with other communication links. For instance, in urban air mobility with high-density LAP traffic, beamforming ensures that energy is precisely concentrated on the intended communication nodes, thereby preventing signal bleed-over into adjacent channels or interfering with other aerial vehicles. 

\par In addition to interference, spectrum scarcity poses another critical challenge in LAE systems. To overcome this challenge, beamforming enhances spectrum utilization by enabling efficient spectrum reuse. Specifically, by tailoring beam directions based on the location of each user or device, multiple users can simultaneously operate on the same frequency band without introducing significant co-channel interference. For example, in live event broadcasting, media LAPs can generate directional beams, which enables multiple LAPs to simultaneously transmit on a shared frequency channel to different ground receivers across a venue, thereby effectively managing co-channel interference.

\subsubsection{Sensing}

Beamforming plays a pivotal role in overcoming key sensing challenges in LAE applications. First, one significant challenge in LAE sensing is maintaining accurate tracking of moving targets, especially those with highly dynamic and unpredictable motion patterns. To address this issue, beamforming dynamically adjusts the beam direction according to the real-time positions of the targets. Specifically, by continuously aligning signal focus with the instantaneous location of each target, beamforming supports consistent tracking precision under mobility. For instance, in LAE logistics scenarios involving LAP swarms, precise and uninterrupted target localization is essential. In such scenarios, adaptive beam control enables each LAP to maintain accurate localization even as target positions fluctuate, ultimately supporting mission reliability.

\par In addition to target tracking, another critical challenge is ensuring robust environmental sensing, particularly when wireless signals are prone to significant attenuation over long distances or when attempting to penetrate obstructions like debris or foliage. Moreover, such attenuation can severely degrade the quality of sensing data. To address this issue, beamforming mitigates attenuation effects by concentrating wireless signal energy toward areas of interest, thereby improving signal strength and increasing the fidelity of received data~\cite{liu2023deployment}. For example, in LAE disaster response missions, concentrated beam scanning enables rapid detection of survivors hidden beneath rubble or obstructive terrain, thus enabling responders to efficiently locate individuals.

\subsection{Existing Challenges}
The effectiveness of conventional beamforming technologies is currently constrained by several unresolved challenges, which are further discussed as follows.

\subsubsection{Scalability Constraints} As wireless communication systems continue to evolve, the scale of deployment in terms of antenna arrays, user density, and spatial coverage is rapidly increasing. In LAE scenarios, where multiple platforms may operate simultaneously, beamforming techniques must be capable of handling large and complex configurations. However, conventional approaches often rely on fixed model assumptions and manually designed optimization procedures, which become increasingly difficult to maintain as the system scales. As a result, this lack of scalability limits their effectiveness in dense and dynamic communication scenarios.

\subsubsection{Real-Time Adaptation}
Although current adaptive beamforming methods can handle certain dynamic adjustments, they often struggle to achieve real-time adaptation in LAE. Tasks such as LAP delivery and low-altitude mapping involve rapidly changing flight paths and wireless channel conditions, which require immediate updates to beam directions. However, conventional approaches rely on infrequent update cycles and limited feedback mechanisms, which lack the responsiveness to track rapid variations in CSI under highly dynamic conditions.

\subsubsection{Multi-Objective Optimization}
Beamforming balances interference reduction, energy efficiency, and communication reliability. However, beamforming often faces stringent environmental and resource constraints. Consequently, overlapping paths and limited spectrum in LAE scenarios make emergency communications and similar high-demand situations increasingly challenging. Therefore, addressing these multi-objective challenges calls for more advanced optimization techniques.

\subsubsection{Antenna Technology Complexity}
Conventional beamforming techniques have significant limitations when faced with advanced antenna technologies like massive multiple-input multiple-output (MIMO), holographic MIMO, and extremely-large MIMO.
The reason is that conventional beamforming designs rely on mathematical and analytical models and struggle to handle the complexity and high demands of these advanced technologies. As a result, conventional beamforming designs cause lower efficiency and limited adaptability, thus highlighting the need for alternative technologies.

 {\color{color_revise}
\subsubsection{Channel Estimation Errors}
In LAE beamforming, channel estimation errors lead to imprecise beam alignment, which degrades signal quality and increases the risk of signal leakage to eavesdroppers. To compensate for this, energy-constrained LAPs must increase transmission power, thus reducing the system lifetime. Moreover, this challenge is exacerbated as the number of collaborative entities increases since individual errors can accumulate and be magnified throughout the network.
}

\par The aforementioned challenges highlight the need for more intelligent and adaptable solutions to improve beamforming in LAE applications. In this context, GenAI technologies provide flexible and efficient methods to address the complexity and dynamic requirements of modern beamforming.

\section{GenAI for Beamforming in Low-Altitude Economy}
\label{Sec:Generative AI for Beamforming in Low-Altitude Economy}

\par In this section, we begin by introducing foundational GenAI models. Next, we discuss how GenAI enhances beamforming and highlight specific advancements in the context of LAE, as presented in Table~\ref{tab:GenAI for bf and lae-bf}.

\subsection{Fundamentals of GenAI}

\subsubsection{Variational Autoencoders (VAEs)} VAEs establish a probabilistic generative framework by explicitly optimizing a lower bound on data likelihood to learn latent representations that capture underlying data structures. Their distinctive advantage lies in naturally quantifying uncertainty through explicit variance modeling, while simultaneously ensuring consistent training convergence, despite typically generating reconstructions with less sharp details. As a result, the inherent uncertainty quantification makes VAEs particularly valuable for channel estimation in rapidly changing LAE conditions, where accurate beam prediction requires modeling the distribution of possible channel states rather than providing single point estimates. In light of these properties, Baur \textit{et al.} \cite{baur2024channel} employed a VAE-based model to learn the complex probability distribution of CSI, thus parameterizing estimators that approach minimum mean square error performance even without perfect CSI during training phases. Moreover, by encoding channel characteristics into a latent space with explicit variance parameters, VAEs provide not only the channel estimates themselves but also confidence measures for those estimates, proving crucial when making communication decisions in variable channel conditions. In addition, the robustness of VAEs stems from this probabilistic generative structure. Specifically, by modeling a distribution over latent variables, VAEs inherently account for uncertainty and variability, thus promoting strong generalization and robust inference from partial, noisy, or degraded input, which is beneficial when sensor data is missing or corrupted.

\subsubsection{Generative Diffusion Models (GDMs)} GDMs learn complex data distributions through a distinctive iterative denoising process. Specifically, GDMs simulate a gradual forward diffusion process that incrementally adds noise to data. Subsequently, the models learn to precisely reverse this process by progressively removing the noise step by step, thereby reconstructing the original data. Therefore, the step-wise denoising mechanism enables GDMs to capture the full transformation from pure noise to structured data, along with underlying statistical structures, thus leading to the generation of exceptionally high-fidelity and diverse samples. Moreover, this remarkable ability to generate accurate and varied data makes GDMs particularly valuable for modeling complex LAE channels, especially when detailed real-world channel measurements are limited or challenging to obtain. In light of these properties, Wu \textit{et al.} \cite{wu2025icdm} applied a GDM-based model to generate realistic MIMO channel samples conditioned on user location, even with limited real-world measurements. In addition, the robustness of GDMs is rooted in this iterative denoising procedure, thereby allowing them to correct arbitrary levels of corruption and achieve resilient reconstruction under strong jamming, complex noise, or signal dropout conditions, all common in low-altitude environments~\cite{Xu2025Semantic}.

\begin{table*}[ht]
    \centering
    \caption{{\color{color_revise}GenAI Models for Beamforming Applications in Low-Altitude Economy}}
    \label{tab:GenAI for bf and lae-bf}
    \begin{tabular}{>{\centering\arraybackslash}p{1.2cm} >{\centering\arraybackslash}m{3.1cm} >{\centering\arraybackslash}m{3.0cm} >{\centering\arraybackslash}m{3.4cm} >{\centering\arraybackslash}m{3.1cm} >{\centering\arraybackslash}m{2.2cm}}
    \toprule
    \textbf{Model} & \textbf{Channel Estimation} & \textbf{Resource Allocation} & \textbf{Communication Security} & \textbf{Energy Consumption} & \textbf{Real-time} \\
    \midrule
    \textbf{VAE} &
        Quantify uncertainty in channel estimation by modeling probability distributions \cite{baur2024channel} &
        Extract features to generate beam and spectrum allocation strategies &
        \multicolumn{1}{c}{-} &
        \multicolumn{1}{c}{-} &
        % \begin{tabular}{@{}c@{}}Text: Moderate \\ Image: Moderate \\ Video: Moderate\end{tabular} \\
        \raisebox{0.5\height}{\begin{tabular}{@{}c@{}}Text: Moderate \\ Image: Moderate \\ Video: Moderate\end{tabular}} \\
    \midrule
    \textbf{GDM} &
        Generate realistic channel samples~\cite{wu2025icdm} and refine noisy observations via iterative denoising~\cite{R3-1-8} &
        \multicolumn{1}{c}{-} &
        Resist eavesdropping interference and mitigate location errors in beamforming \cite{zhao2025generative} &
        Iteratively refine beam patterns for energy-efficient power distribution &
        % \begin{tabular}{@{}c@{}}Text: Fair \\ Image: Fair \\ Video: Fair\end{tabular} \\
        \raisebox{0.5\height}{\begin{tabular}{@{}c@{}}Text: Fair \\ Image: Fair \\ Video: Fair\end{tabular}} \\
    \midrule
    \textbf{GAN} &
        Generate realistic samples to fill information gaps &
        \multicolumn{1}{c}{-} &
        Suppress complex interference patterns through adversarial denoising \cite{tang2023wireless} &
        Simulate low-power beam patterns &
        % \begin{tabular}{@{}c@{}}Text: Excellent \\ Image: Excellent \\ Video: Good\end{tabular} \\
        \raisebox{0.5\height}{\begin{tabular}{@{}c@{}}Text: Excellent \\ Image: Excellent \\ Video: Good\end{tabular}} \\
    \midrule
    \textbf{Transformer} &
        Use a sequence of previous
        channel states to predict future channel parameter~\cite{R3-1-9} &
        Capture long-range dependencies and context for resource allocation &
        Enhance secrecy via beamforming and spatiotemporal modeling \cite{zhao2025generative} &
        Optimize energy consumption by capturing environmental dependencies &
        % \begin{tabular}{@{}c@{}}Text: Good \\ Image: Good \\ Video: Moderate\end{tabular} \\
        \raisebox{0.5\height}{\begin{tabular}{@{}c@{}}Text: Good \\ Image: Good \\ Video: Moderate\end{tabular}} \\
        
    \bottomrule
    \end{tabular}
\end{table*}

\subsubsection{Generative Adversarial Networks (GANs)} GANs learn data distributions through an adversarial training framework, where a generator and a discriminator are trained in competition to implicitly capture underlying data structures. Crucially, the key strength of GANs lies in their ability to learn complex data patterns without needing explicit mathematical modeling of the data distribution. Therefore, the capability is particularly advantageous for interference suppression in LAE, which is characterized by non-stationary and complex interference patterns. Moreover, by training on clean signal data, GANs can learn what constitutes a clean signal and then use this knowledge to effectively remove interference from corrupted signals. As a result, GAN-based architectures enable robust interference mitigation even when the interference characteristics are unknown or constantly changing. For example, in~\cite{tang2023wireless}, the authors proposed a GAN-based model for wireless signal denoising where the generator reconstructed clean signals from noisy inputs and the discriminator ensured that the outputs were indistinguishable from real clean signals, significantly improving clarity under complex noise and interference. In addition, the robustness of GANs arises from these adversarial training dynamics. Specifically, continuous competition forces the generator to approximate data distributions, including rare or extreme variants, which promotes strong resilience to distributional shifts and adversarial perturbations, frequently occurring in dynamic wireless environments.

\subsubsection{Transformers} Transformers leverage global context through self-attention mechanisms, which generate content by modeling the conditional probability of each element in the sequence. A key advantage of self-attention lies in its ability to directly capture long-range dependencies and relationships within sequential data, rather than relying on fixed-size windows or recurrent connections. Consequently, the inherent capability for modeling comprehensive spatiotemporal relationships makes Transformers particularly valuable for tasks requiring strong temporal and spatial awareness. For example, in \cite{wang2021deep}, the authors embedded dynamic network states into structured tokens and applied a Transformer-based model to model multi-scale spatiotemporal dependencies. The resulting representations effectively captured time-varying channel dynamics and user preferences, thereby enhancing the exploration capability of the actor-critic network in deep reinforcement learning (DRL). In addition, the robustness of Transformers derives from the self-attention mechanism and their capacity to process non-sequential, variable-length, and partially corrupted input. Specifically, self-attention enables the model to dynamically reweight input tokens based on contextual relevance, thus allowing the model to suppress noisy features and infer missing information by referencing global context. Consequently, the robustness of Transformers makes them particularly well suited for handling irregular input streams, fluctuating channel quality, and asynchronous data transmissions in LAE scenarios.

\subsection{GenAI for Beamforming in Low-Altitude Economy}

% \par GenAI elevates beamforming in LAE by dynamically adapting to complex environmental uncertainties and collaborative network demands. This section demonstrates its capabilities in mission-critical tasks and resilient performance enhancement.

\par GenAI offers a holistic enhancement to beamforming in the LAE, fundamentally improving both core communication tasks and overall system performance.

\subsubsection{Communication Tasks}
\par The dynamic nature of LAE communication demands highly agile and intelligent management. To this end, GenAI provides a powerful framework for robustly executing critical communication tasks.

\begin{itemize}
    \item \textbf{Channel Estimation:} 
     {\color{color_revise} Compared to conventional communication scenarios, there is an increased demand for channel estimation of LAE due to the complex and highly dynamic environment. Specifically, some factors such as adverse weather conditions, obstruction by buildings and the rapid mobility of LAPs often lead to signal attenuation and pronounced multipath propagation effects, thus posing substantial challenges to accurate channel estimation. In this case, GenAI models can offer a full-cycle solution by addressing different stages of the channel information processing. The initial stage involves incomplete historical channel information reconstruction, where GANs can be used to generate realistic samples to fill information gaps. Moreover, for the current channel state analysis, VAEs can provide a probabilistic assessment by modeling the channel distribution. Furthermore, GDMs can refine noisy observations through iterative reverse diffusion. For instance, in~\cite{R3-1-8}, the authors proposed a GDM-based model that frames channel estimation as a denoising problem and recovers the true channel by reversing a noise-adding process. In addition, for future channel information prediction, Transformers can leverage self-attention to capture temporal correlations within sequential data. For example, Ju \textit{et al.} \cite{R3-1-9} developed a Transformer-based framework that uses a sequence of previous channel states to predict future channel parameters and thereby mitigate channel aging.
    }

    \item \textbf{Resource Allocation:} Compared to conventional communication environments, efficient spectrum and power resource allocation in LAE scenarios is more challenging due to dense node deployments, rapid network topology changes caused by node mobility, and intense competition for limited resources. However, conventional beamforming approaches often fail to quickly adjust beam configurations in response to these complex factors, thus leading to inefficient resource management. In contrast, GenAI-based beamforming methods explicitly link beamforming optimization with resource allocation, enabling simultaneous and real-time adjustment of beam directions and spectrum assignments to address resource conflicts and enhance utilization efficiency. For instance, a VAE-based framework can encode network topology variations and resource competition conditions into concise latent features. Subsequently, the decoder directly generates optimized beamforming parameters and adaptive spectrum allocation schemes. This integrated approach effectively mitigates interference, resolves resource competition conflicts, and improves overall spectrum and power utilization efficiency in highly dynamic LAE scenarios. 
\end{itemize}

\subsubsection{Performance Metrics}
 \par LAE demands adaptive security and energy resilience against evolving threats. To address this demand, GenAI shows promising progress, which can be described as follows.

\begin{itemize}
    \item \textbf {Communication Security:}
     Compared to the conventional communication scenarios, communication security in LAE is particularly critical. On the one hand, different tasks may rely on different networks or frequency bands, which means that the LAPs frequently need to switch between multiple communication networks and frequency bands. As a result, this increases the vulnerability of communication links to network attacks or malicious interference. On the other hand, LAE includes tasks in critical areas such as logistics and emergency rescue, where the reliability of communication security directly impacts the efficiency and safety of these services. However, conventional beamforming algorithms rely solely on signal strength to optimize beam direction, which limits their ability to handle dynamic malicious interference. In contrast, GenAI models excel at analyzing latent data distributions and dynamically optimizing anti-jamming strategies under uncertain conditions. For example, in \cite{zhao2025generative}, the authors combined a diffusion-based optimization framework with a mixture of expert (MoE)-Transformer network, and leveraged multi-head attention to model spatiotemporal channel dependencies. Moreover, this method dynamically optimizes the three-dimensional beamforming to resist eavesdropping interference, mitigate location errors, and address channel uncertainties, thus enhancing the anti-jamming resilience and secrecy performance in the mission-critical low-altitude communication systems.

    \item \textbf {Energy Consumption:} In LAE scenarios, the frequent beam redirection by LAPs, interference suppression between high-density nodes, and multi-network handovers significantly increase LAP energy consumption compared to conventional communication environments. However, conventional beamforming algorithms rely on preset power distribution strategies, which hinder the ability to balance communication quality and energy efficiency in dynamic topology and channel fluctuation scenarios. GenAI can learn the relationship between environmental features and energy consumption online, autonomously generating beam control strategies that optimize energy efficiency and overcome the limitations of traditional static approaches. For instance, a beam control system based on GAN can employ a generator to simulate low-power beam patterns and dynamically optimize beam width and radiation power. Meanwhile, the discriminator can assess signal integrity based on real-time channel states, which ensures a reduction in energy consumption without sacrificing performance.
\end{itemize}

%
% Figure of Case_Study
%
\begin{figure*}[!hbt] 
	\centering
	\includegraphics[width =0.95\textwidth]{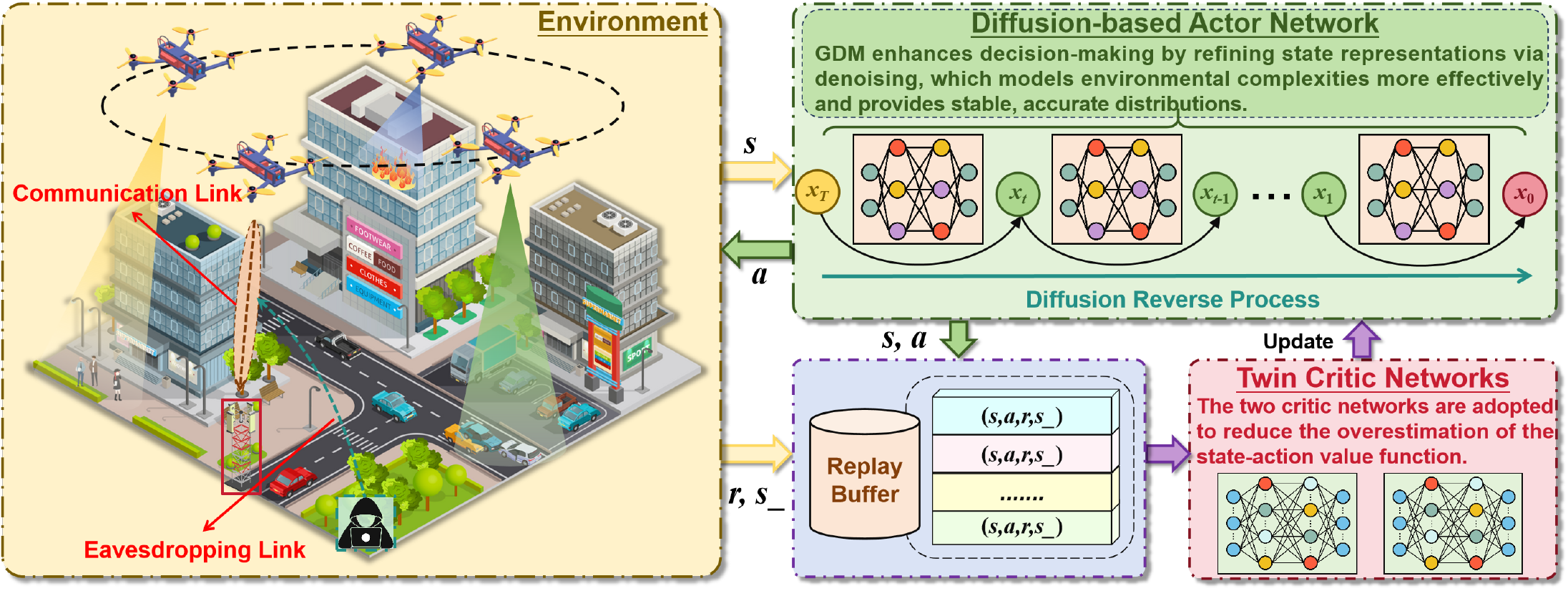}
	\caption{The workflow of the GDMTD3 algorithm includes environmental observation, action generation via a diffusion-based actor network, evaluation by twin critic networks, and policy updates supported by a replay buffer.}
	\label{Case_Study_Workflow}
\end{figure*}

\subsection{Lessons Learned}

\par In this section, we explored the application of GenAI models in beamforming, particularly within the context of LAE. From these insights, we derive the following key lessons.

\begin{itemize}    
    \item GenAI models fundamentally enhance the adaptability, inherent robustness, and real-time responsiveness in beamforming processes, particularly in complex, dynamic LAE environments. Specifically, the capacity to learn underlying data distributions, quantify uncertainty, and efficiently infer from complex inputs enables rapid adaptation to changing channel conditions and resilient data reconstruction even under severe signal degradation. Moreover, the operational agility is vital for maintaining high-quality, continuous communication and precise tracking across diverse data types, including text, images, and video, thus ensuring robust performance despite environmental uncertainties and mobility constraints.
    {\color{color_revise}
    \item GenAI significantly improves channel estimation in LAE environments. Specifically, GenAI models can offer a full-cycle solution that addresses different stages of the estimation process. In addition, GenAI models also fundamentally redefine resource allocation. In particular, GenAI-enhanced beamforming methods can simultaneously adjust signal directions and spectrum assignments.}
    \item For communication security, GenAI models offer significant advancements in LAE. Specifically, GenAI models adaptively optimize anti-jamming strategies against dynamic threats, which are crucial for critical missions~\cite{Li2025Securingthesky}. Moreover, GenAI provides solutions for energy consumption. In particular, GenAI-based models learn environmental features online to generate optimized, energy-efficient beam control strategies, thereby maintaining performance while reducing power use.
\end{itemize}

\section{Case Study: GenAI for collaborative beamforming in Low-Altitude Economy}
\label{Sec:Case Study}

\par In this section, we examine a GenAI-based approach for collaborative UAV beamforming in LAE scenarios. First, we discuss the motivation behind this approach and provide an overview of the system design. Next, we introduce the proposed algorithm. Finally, we evaluate how this method enhances the communication efficiency and security.

\subsection{Motivation and System Design}

\par In LAE scenarios, security risks are rising due to complex terrains and high-density LAP deployments, often leading to communication vulnerabilities where traditional encryption and anti-jamming methods struggle under resource constraints. Therefore, high secrecy rates are crucial to prevent eavesdropping on sensitive missions in open airspace. Additionally, minimizing energy consumption is vital for long-term operations of LAPs, given their limited onboard batteries and high energy demands. Moreover, a tradeoff exists between secrecy rate and energy consumption. Specifically, boosting security might mean LAPs fly longer or more complex paths, increasing energy use. Conversely, low battery levels could prevent energy intensive security maneuvers. As a result, joint optimization of both is essential for a practical balance. 

\par As the precise synchronization and high-precision localization technologies for distributed aerial platforms have matured, collaborative beamforming has emerged as an effective means of achieving long-distance secure communications. Therefore, this case study utilizes UAVs, key components of LAPs, to implement collaborative aerial beamforming for secure remote communications. Specifically, multiple UAVs form a virtual antenna array to enhance the signal directionality and strength towards the remote base station, while optimizing excitation current weights and UAV positions to ensure secure and energy-efficient data transmission. The proposed optimization framework combines the GDM model with DRL to improve both security and energy efficiency.

\subsection{Simulation Setup} 
\par In this section, we detail the computational environment and environmental parameters used to evaluate our proposed framework.

\begin{itemize}
    \item \textbf{Simulation Platform:} Our experiments are conducted using a computing setup that included an NVIDIA GeForce RTX 3090 GPU with 24 GB of memory and a 13th Gen Intel(R) Core(TM) i9-13900K 32-core processor with 128 GB of RAM. The operating system on the workstation is Ubuntu 22.04.3 LTS. For our deep learning computations, we use PyTorch 2.6.0, along with the CUDA 12.4.
    
    \item \textbf{Environmental Details:} In this study, we consider a UAV swarm consisting of 4 units, each equipped with a transmit power of 0.1 W. Moreover, the swarm is randomly distributed within a 40 m × 40 m area. To simulate potential security threats, a mobile eavesdropper is introduced, which follows the Gauss-Markov mobility model with an average speed of 5.0 m/s, a correlation coefficient of 0.1, and a variance of 1.0.
\end{itemize}

\subsection{Proposed Algorithm}

\par Considering the challenges posed by the dynamic and uncertain nature of LAE environments, common DRL algorithms often struggle to achieve stable and proper performance. To address these limitations, this study proposes the GDM-enhanced twin delayed deep deterministic policy gradient (GDMTD3) algorithm, which integrates a GDM into the twin delayed deep deterministic policy gradient (TD3) framework to enhance the comprehension of environmental dynamics. This integration allows the algorithm to optimize UAV swarm actions by modeling the problem as a Markov decision process. Specifically, the state is defined by the positions of the UAVs and the estimated location of the eavesdropper. Based on this, the action consists of selecting excitation current weights and adjusting UAV positions. Furthermore, the reward function combines the secrecy rate and energy consumption, while also imposing penalties for violations such as exceeding speed limits or UAV collisions.

\par Fig. \ref{Case_Study_Workflow} shows the framework of the proposed GDMTD3 algorithm. Specifically, the workflow begins with the system observing the current environment, which includes the positions of the UAVs and the estimated location of the eavesdropper. Based on this information, the diffusion-based actor network processes the state and generates the candidate actions. Specifically, GDM replaces the standard actor network in the TD3 framework, where it learns the probability distribution of optimal actions and applies a reverse diffusion process. This allows the algorithm to refine noisy inputs into high-quality actions that are better tailored to the current environment. Subsequently, these actions are evaluated by the twin critic networks, which compute the corresponding secrecy rate and energy consumption. In turn, the evaluations provide crucial feedback that guides the actor policy, thus helping it make optimal trade-offs between security and efficiency. Furthermore, the replay buffer stores historical state-action-reward tuples, which are then sampled to iteratively update both the actor and critic networks. This iterative process ensures policy convergence to actions that enhance communication security, minimize energy consumption, and maintain stability in training. Through this closed-loop process, the algorithm dynamically adapts to environmental changes, thereby achieving robust performance in both security and energy efficiency.

\par In addition to the above, the proposed system also possesses significant resilience and broad applicability across diverse environments. In particular, the GenAI-enabled framework integrated into system design adapts to varying device mobility and heterogeneous antenna configurations through dynamic learning capability. Moreover, decoupled offline training ensures compatibility with resource-constrained edge deployments. Furthermore, inherent robustness allows GDMTD3 to dynamically handle abnormal situations by perceiving such occurrences as environmental changes and then adjust policies accordingly. For example, if a UAV in the swarm experiences a fault, the system immediately perceives this as an environmental change. Consequently, GDMTD3 algorithm then dynamically adjusts the communication parameters, such as re-routing data traffic through other functional UAVs or reallocating tasks to more reliable nodes, to maintain robust data links and ensure reliable operation despite the failure of individual UAVs.

% figure of simulation
\begin{figure}[htbp]
    \centering
    \includegraphics[width=\linewidth]{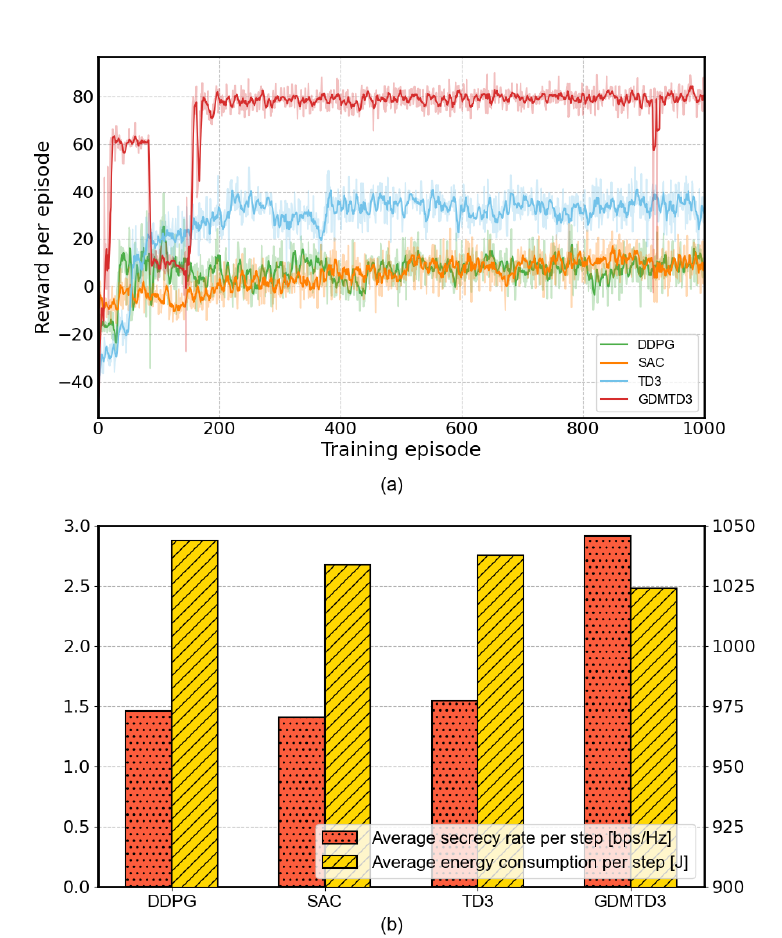} 
    \caption{Performance evaluation of the GDMTD3 algorithm compared with DDPG, SAC, and TD3. (a) Average reward per episode. (b) Average secrecy rate and energy consumption per step.}
    \label{fig:performance_evaluation}
\end{figure}

\subsection{Performance Evaluation}
\par To validate the effectiveness of the proposed GDMTD3 algorithm, extensive simulations were conducted and compared with baseline algorithms, including deep deterministic policy gradient (DDPG), soft actor-critic (SAC), and standard TD3.

\par As shown in Fig. \ref{fig:performance_evaluation}(a), the GDMTD3 algorithm achieved higher and more stable rewards across 1000 training iterations compared to the baseline methods. This demonstrates its ability to learn optimal policies effectively and adapt to dynamic environments, while the baseline algorithms exhibited slower convergence and greater variability. In terms of optimization objectives, Fig. \ref{fig:performance_evaluation}(b) shows that the proposed GDMTD3 achieved the highest secrecy rate with the lowest energy consumption per step, effectively balancing the security and efficiency. In contrast, SAC and TD3 showed moderate performance in one metric but failed to achieve the optimal trade-off.

\section{Conclusion and Future Directions}
\label{Sec:Conclusion and Future Directions}

\par In this article, we have introduced the review of applying GenAI models in beamforming for LAE. We have first analyzed the communication requirements in LAE and the limitations of conventional beamforming techniques. Next, we have explored the applications of GenAI models in two aspects, namely beam management and beam tracking, and highlighted its potential to enhance the communication performance. Subsequently, we have presented a case study to verify the effectiveness of the proposed GDM-based collaborative beamforming method in improving the communication security and energy efficiency for LAE. In addition, three other directions for future research in GenAI techniques for beamforming in LAE scenarios are outlined below. 

\subsection{Semantic-Driven Beamforming}
As communication demands in LAE scenarios shift from traditional data transmission to semantic information delivery, existing beamforming techniques struggle to meet precise requirements. In response, integrating beamforming with semantic communication frameworks is a promising direction. This approach dynamically adjusts beam directions and power allocation based on semantic task demands. As a result, it can better support evolving communication needs.

\subsection{Long-Range High-Power Wireless Power Transfer-Driven Beamforming}
Unlike short-range wireless power transfer (WPT) targeting low-power devices and sensors, LAPs require long-range, high-power WPT for reliable and sustainable operation. In this context, it is necessary to realize long-range high-power WPT with beamforming, such as laser and radio frequency (RF)-based power beamforming. Moreover, integrating space-based solar power with such advanced WPT solutions can address energy challenges, ultimately supporting the development of 6G systems across various applications in LAE.

\subsection{Integrated Sensing and Communications-Driven Beamforming}
Multiple independent sensing beams are necessary to detect separate targets in integrated sensing and communications (ISAC) scenarios anticipated for 6G systems, particularly in LAE applications. This is due to the fact that near-field beam focusing with large-scale MIMO technology concentrates transmit signal energy in specific directions and distances, which differs from the far-field ISAC systems. Consequently, future research on ISAC-driven beamforming (i.e., beam focusing) for LAE applications should prioritize the use of multiple narrow beams to optimize sensing performance and ensure efficient operation of LAPs.

\ifCLASSOPTIONcaptionsoff
\newpage
\fi

% \bibliographystyle{IEEEtran}
% \bibliography{references.bib}  
\bibliographystyle{ieeetr}

% % 参考文献部分
% {\tiny  % 设置参考文献部分字体为小号
% \printbibliography
% }
  
\vfill

\end{document}